\documentclass[aps,pra,twocolumn,amsmath,amssymb,superscriptaddress]{revtex4-1}
\usepackage{epsfig}
\usepackage{epstopdf}
\usepackage{bbold}
\usepackage{color}

\newcommand{\ket}[1]{|#1\rangle}
\newcommand{\bra}[1]{\langle #1|}
\newcommand{\ip}[2]{\langle #1|#2 \rangle}

\begin{document}

%\preprint{}

\title{Higher-dimensional orbital angular momentum based quantum key distribution with mutually unbiased bases}



%\author{*}
%\affiliation{*}


\author{Mhlambululi Mafu}
\affiliation{School of Chemistry and Physics, University of KwaZulu-Natal, Private Bag X54001, Durban 4000, South Africa}
\author{Angela Dudley}
\affiliation{CSIR National Laser Centre, P.O. Box 395, Pretoria 0001, South Africa}
\author{Sandeep Goyal}
\affiliation{School of Chemistry and Physics, University of KwaZulu-Natal, Private Bag X54001, Durban 4000, South Africa}
\author{Daniel Giovannini}
\affiliation{School of Physics and Astronomy, SUPA, University of Glasgow, Glasgow, UK}
\author{Melanie McLaren}
\affiliation{CSIR National Laser Centre, P.O. Box 395, Pretoria 0001, South Africa}
\author{Miles J. Padgett}
\affiliation{School of Physics and Astronomy, SUPA, University of Glasgow, Glasgow, UK}
\author{Thomas Konrad}
\affiliation{School of Chemistry and Physics, University of KwaZulu-Natal, Private Bag X54001, Durban 4000, South Africa}
\affiliation{National Institute for Theoretical Physics (NITheP), University of KwaZulu-Natal, Private bag X54001, Durban 4000, South Africa}
\author{Francesco Petruccione}
\affiliation{School of Chemistry and Physics, University of KwaZulu-Natal, Private Bag X54001, Durban 4000, South Africa}
\author{Norbert L\"{u}tkenhaus}
\affiliation{Institute for Quantum Computing \& Department for Physics and Astronomy, University of Waterloo, 200 University Avenue West, N2L 3G1,
Waterloo, Ontario, Canada}
\author{Andrew Forbes}
\affiliation{CSIR National Laser Centre, P.O. Box 395, Pretoria 0001, South Africa}

%\thanks{}
%\altaffiliation[Also at ]{Laser Research Institute, University of Stellenbosch, %Stellenbosch 7602, South Africa}

%\altaffiliation[Also at ]{Laser Research Institute, University of Stellenbosch, %Stellenbosch 7602, South Africa}


\date{\today}

\begin{abstract}
We present an experimental study of higher-dimensional quantum key distribution protocols based on mutually unbiased bases, implemented by means of photons carrying  orbital angular momentum. We perform ($d+1$) mutually unbiased measurements in a classical prepare and measure scheme and on a pair of entangled photons for dimensions ranging from $d = 2$ to 5. In our analysis, we pay attention to the detection efficiency and photon pair creation probability.  As security measures, we determine from experimental data the average error rate, the mutual information shared between the sender and receiver and the secret key generation rate per photon. We demonstrate that increasing the dimension leads to an increased information capacity as well as higher key generation rates per photon up to a dimension of $d = 4$. %some point and then it goes down. 
\end{abstract}

%\keywords{}

\maketitle

\section{Introduction}
Quantum key distribution (QKD) establishes a secure key between two parties, Alice and Bob, in which they can encode a secret message \cite{RevModPhys.74.145, RevModPhys.81.1301, C.H.Bennet}. Protocols for QKD are classified as either prepare and measure (P\&M) schemes or entanglement-based (EB) schemes. Examples of P\&M schemes are BB84 \cite{C.H.Bennet}, B92 \cite{PhysRevLett.68.3121}, six-state \cite{PhysRevLett.81.3018} and SARG04 \cite{PhysRevLett.92.057901}. However, P\&M schemes such as the E91 protocol \cite{ekert1991quantum} in general can be translated into EB schemes. 

Mutually unbiased bases (MUBs) \cite{schwinger1960unitary, ivonovic1981geometrical, wootters1989optimal} have found many applications, for example in quantum state tomography \cite{wootters1989optimal, PhysRevLett.105.030406, PhysRevA.83.052332, giovanni2013} and quantum error correction codes \cite{calderbank1997quantum, PhysRevA.54.1862} and also appear useful  in QKD protocols. This is because projective measurements in one basis provides no knowledge of the state in any of the other bases \cite{C.H.Bennet, barnett2009, durt2010}. Therefore if an eavesdropper measures in the incorrect basis he/she will obtain no meaningful information but instead introduce a disturbance in the system, resulting in its detection. The simplest example of MUBs of dimension $d = 2$ are the horizontal/vertical, diagonal/anti-diagonal, and left-/right-handed polarization bases as they are unbiased with respect to each other, forming a set of three MUBs. Although MUBs offer security against eavesdropping, encoding states in the polarization degree of freedom only allows a maximum of one bit of information transmitted per photon which results in a limited key generation rate. Since systems with higher-dimensional Hilbert space can store more information per carrier, the question arises whether QKD protocols using higher-dimensional MUBs also result in higher generation rates of secure key bits; indeed, such protocols can be expected to be more robust in terms of abstract noise measures \cite{PhysRevA.82.030301, PhysRevA.85.052310}. Their actual performance in terms of secure key rate however, depends on whether the amount of noise in higher-dimensional implementations grows  faster with increasing dimension than their robustness against noise. The present article addresses this question for  implementations using the orbital angular momentum (OAM) of photons. Beams that carry OAM have an azimuthal angular dependence of exp$(i\ell\theta)$ \cite{Allen1992} where $\ell$ is the azimuthal index and $\theta$ is the azimuthal angle.
%Since higher-dimensions offer an improvement in secure information capacity \cite{leach2012}, encoding MUBs in the OAM degree of freedom is known to lead to higher key generation rates per photon \cite{wiesniak2011entanglement}, as well as higher-dimensional MUBs  offer increased security \cite{PhysRevA.82.030301, PhysRevA.85.052310}. 
It has been shown theoretically that MUBs for higher-dimensional OAM states can be used to encode bits of information in alignment with the BB84 protocol \cite{bechmann2000quantum, groblacher2006experimental, PhysRevA.78.012344, PhysRevLett.88.127902}. A standard P\&M implementation of a generalized BB84 protocol, relying on 11 OAM states and superpositions of these 11 OAM states, has previously been performed \cite{BoydQKD}, using 2 of the 12 available MUBs.

In this paper we experimentally investigate an entanglement-based scheme for QKD encoded in complete sets of higher-dimensional MUBs, which we first verify with a classically simulated P\&M scheme.  
%Although we only focus on the quantum phase of the protocol and do not execute the classical communication channel, \textcolor{blue}{we show that the quantum phase is good enough such that one can execute the whole protocol.}  
We implement our protocol with MUBs encoded in OAM states and present values for the corresponding average error rates, classical Shannon information and secret key rates. As with all OAM protocols, our QKD protocol uses filter measurements that project onto one MUB element at a time; we provide the connection between these protocols to the established theory for protocols using full MUB measurements. To achieve this, we prove that the detection efficiency depends only on the basis choice and not on the elements within a basis, otherwise the security parameters of the protocol cannot be evaluated. This allows us to map our protocol to the key rates, thus arriving at the standard MUB protocol. 
%In addition, we show how to relate the security analysis based on measurements of entire bases to the analysis for filter measurements which select only single basis elements. The key to  relate the former to the latter is to show that the efficiencies to detect single basis states belonging to the same basis are identical. 
%We show how one migrates from a filter measurement based protocol to a theory proof that uses the entire basis measurement. To achieve this, we prove that the detection efficiency depends only on the basis choice and not on the elements within a basis, otherwise the security parameters of the protocol can not be evaluated. This allows us to map our protocol to the key rates, thus arriving at the standard MUBs protocol. 
By increasing the dimension $d$, we obtain an increase in the secret key rate which has been theoretically observed in recent papers \cite{PhysRevA.82.030301, PhysRevA.85.052310}, resulting in higher key generation rates for dimension $d=4$. Similarly the Shannon mutual information increases, demonstrating an improvement in the information capacity.  

\section{Choice of Mutually unbiased bases}
Two orthonormal bases $\mathcal{M}_1 = \{\ket{\phi_{(1,i)}}, i = 0,1,\cdots , d-1\}$ and $\mathcal{M}_2 = \{\ket{\phi_{(2,j)}}, j = 0,1,\cdots , d-1\}$ of a $d$-dimensional Hilbert space $\mathcal{H}_d$ are said to be mutually unbiased if, and only if,   all pairs of basis vectors $\ket{\phi_{(1,i)}}$ and $\ket{\phi_{(2,j)}}$ satisfy
\begin{align}
|\ip{\phi_{(1,i)}}{\phi_{(2,j)}}|^2 & = \frac{1}{d}.
\end{align}

Physically, this means that for a system prepared in the basis $\mathcal{M}_1$ and measured with respect to basis $\mathcal{M}_2$, all outcomes are equally probable. This property of mutually unbiased bases makes them important for QKD protocols. 

Mutually unbiased bases were introduced by Schwinger \cite{schwinger1960unitary} in 1960 as optimum incompatible measurement bases. In 1981, Ivonovic showed their application in quantum state discrimination \cite{ivonovic1981geometrical}. Later Wootters and Fields \cite{wootters1989optimal} gave a constructive  proof that there exist complete sets of MUBs for prime power dimensions and proved that for any dimension $d$ there are not more than $d+1$ MUBs within any particular set of MUBs.

The smallest prime dimension is $2$, and for that an example of a complete set of MUBs  consists of the eigenstates of the three Pauli spin operators $\sigma_z,\sigma_x, \sigma_y$, i.e,
\begin{align}
\{\ket{0},\ket{1}\};~\\
 \left\{\frac{1}{\sqrt{2}}\left(\ket{0} + \ket{1}\right),\frac{1}{\sqrt{2}}\left(\ket{0}-\ket{1}\right)\right\};~\\
 \left\{\frac{1}{\sqrt{2}}\left(\ket{0} +i \ket{1}\right),\frac{1}{\sqrt{2}}\left(\ket{0}-i\ket{1}\right)\right\}.
\end{align}
Pauli operators can be generalized to higher dimension, known as the Weyl operators. These are unitary operators of the form $X^kZ^l$ for $k,l \in \{0,1,\cdots,d-1\}$. The operator $Z$  is diagonal  in the standard basis $\{\ket{0},\ket{1},\cdots ,\ket{d-1}\}$:
\begin{align}
Z&= \sum_{i=0}^{d-1}\omega^i\ket{i}\bra{i},
\end{align}
with $\omega = \exp(i2\pi/d)$ whereas, the operator $X$ reads:
\begin{align}
X &= \sum_{i=0}^{d-1} \ket{i+1~\mbox{mod}~d}\bra{i}.
\end{align}
The eigenbases belonging to the different operators in the set  $\{ Z, XZ^l|l\in\{0,1,\cdots, d-1\}\}$ form a complete set of MUBs for any prime number $d$ as the dimension of the underlying Hilbert space. For $d=2$, the operator $X$ is identical with the Pauli operator $\sigma_x$ and the operator $Z$ is given by the Pauli operator  $\sigma_z$.

In the present study of MUB based QKD a complete set of MUBs is implemented following the recipe above by means of photons carrying OAM. The MUBs are obtained by assuming that the standard basis (eigenbasis of the operator \textit{Z}) is realized by  single-photon states which correspond to an elementary excitation of  Laguerre-Gauss modes (LG$_\ell $) carrying  OAM value $l\hbar$. For $d=2$ we employ the LG$_\ell$ modes with  $\ell=\pm1$ to generate the standard basis.

\begin{figure}[htbp]
\centerline{\includegraphics[scale=1.0]{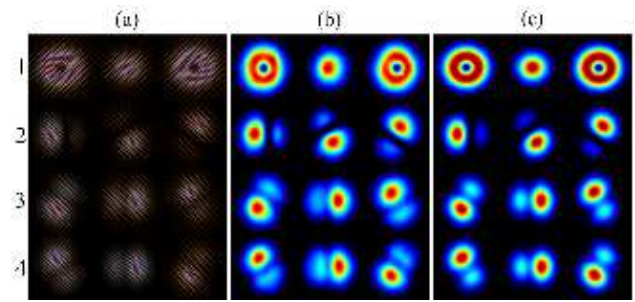}}
\caption{The states for each of the 4 MUBs for $d = 3$. The images on the left represent the measurement filters (or holograms) for each of the 12 states. The images in the middle and on the right contain the corresponding experimentally produced and theoretically calculated intensity profiles of the LG$_\ell$ modes produced by each hologram.}
\label{fig:example}
\end{figure}

For $d=3$, our choice of the  standard basis corresponds to LG$_\ell$ modes  with OAM values $\ell=-1,\,0,\,1$:
\begin{align}
\left\{\ket{-1}\equiv \left(\begin{array}{c}1\\0\\0\end{array}\right),
\ket{0}\equiv \left(\begin{array}{c}0\\1\\0\end{array}\right),
\ket{1}\equiv \left(\begin{array}{c}0\\0\\1\end{array}\right)\right\};\label{single}
\end{align}
The remaining three bases are given in matrix notation with respect to the standard basis as:
\begin{equation}\label{matrices}
\frac{1}{\sqrt{3}}
\begin{pmatrix} 1 & 1 & 1 \\ 1 & \omega & \omega^2 \\ 1 & \omega^2 & \omega \end{pmatrix}, \frac{1}{\sqrt{3}}
\begin{pmatrix} 1 & 1 & \omega \\ 1 & \omega & 1 \\ \omega & 1 & 1 \end{pmatrix}, \frac{1}{\sqrt{3}}
\begin{pmatrix} 1 & 1 & \omega^2 \\ 1 & \omega^2 & 1 \\ \omega^2 & 1 & 1 \end{pmatrix}.
\end{equation} 
Here each matrix represents a complete orthonormal basis with its columns reflecting the basis vectors. In general, for prime dimension $d$,  the standard basis consists of  $d$ LG$_\ell$ modes, while  the remaining  $d$ bases pertain to superpositions of the LG$_\ell$ modes. Examples of the LG$_\ell$ modes and their superpositions are given in Fig.~(\ref{fig:example}), which contains images of the measurement holograms and their corresponding intensity profiles.

\section{Filter based MUB QKD protocol}
We will now describe how our QKD protocol which is based on filter measurements operates. In both scenarios, Alice (SLM A) prepares her mode in a state chosen randomly from one of the ($d + 1$) bases, while Bob (SLM B) performs a measurement on his mode by randomly selecting a state in one of the ($d + 1$) bases chosen out of $d(d + 1)$ different basis settings but biased towards one basis. Each party then announces from which basis the filter measurement was chosen (sifting) and keeps measurements if they all arrived in the same basis. They later make announcements as to whether photon coincidences occurred (post-selection). A  coincidence event represents a conclusive result, otherwise it becomes inconclusive. This is followed by parameter estimation (error rate in the remaining data), error correction and privacy amplification. The announcement  step allows our  filter measurement based QKD protocol to be mapped back to the original protocol which uses full MUB measurements. 
\section{Average error rate and secret key rate}

In standard EB QKD protocols both parties perform measurements on the states that they receive, followed by a public announcement of their measurement basis. The two parties then compare a small portion of their measurements in order to obtain an estimate of the average error rate. This quantifies the error in the QKD protocol resulting from all sources of noise, such as noise in the transmission channel and errors in the measurements. Moreover, the noise could also be caused by an eavesdropper. 
%The average error rate (or quantum bit error rate) is given by the probability that Alice projects onto a different element of a basis than Bob, averaged over the set $\mathcal{L}$ of MUBs used in the QKD protocol. For a given choice of basis $\mathcal{B}_\beta$ in $\mathcal{L}$, the corresponding quantum bit error rate $Q^{\beta}$ is expressed by 
%\begin{equation}
%Q^\beta=\sum_{\stackrel{k,k}{k'\neq k}}\textrm{Tr}\left[|\phi_{(\beta,k)}^*\rangle \langle\phi_{(\beta,k)}^*|\otimes |\phi_{(\beta,k')}\rangle \langle\phi_{(\beta,k')}|\rho_{AB}\right].
%\end{equation}
%The total quantum bit error rate is the total error obtained as an average over the different  bases from the set $\mathcal{L}$ \cite{PhysRevA.85.052310}
%\begin{equation}\label{9}
%Q=\frac{1}{\mathcal{L}}\sum_{\beta\in\mathcal{L}}Q^\beta.
%\end{equation}
%As we use a complete set of MUBs, the number of bases, $|\mathcal{L}|$, is here equal to $d+1$. 
The error rate refers to the probability that Alice sends the state $|\phi_{(\beta,k)}\rangle$, while Bob receives an orthogonal state $|\phi_{(\beta,k')}\rangle$. Given the MUB $\beta$, the corresponding average error rate in each basis $Q^{\beta}$, is expressed as  
\begin{equation}
Q^\beta=\sum_{\stackrel{k,k}{k'\neq k}}tr\left[|\phi_{(\beta,k)}^*\rangle \langle\phi_{(\beta,k)}^*|\otimes |\phi_{(\beta,k')}\rangle \langle\phi_{(\beta,k')}|\rho_{AB}\right].
\end{equation}
The total average error rate is the total error obtained as an average over the different MUBs, $\mathcal{L}$ \cite{PhysRevA.85.052310} and is defined as
\begin{equation}\label{9}
Q=\frac{1}{\mathcal{L}}\sum_{\beta\in\mathcal{L}}Q^\beta.
\end{equation}
We use the full set of available MUBs, therefore $\mathcal{L}=d+1$.
%There is a maximal admissible error rate $Q_{\mbox{max}}$  beyond which no secure key bits can be produced, since the corresponding noise might be due to eavesdropping.  $Q_{\mbox{max}}$ thus is a measure of the robustness of the protocol against noise.

Another important figure of merit for the perfomance of a QKD scheme is the secret key rate. It is given by the amount of information that one can send securely in a photonic QKD scheme. It equals the number of key bits per photon measured by both parties in the same basis that can be generated securely. The maximum secret key rate that one can achieve is log$_{2}d$ for a $d$ level system but is limited by an adversarial attack by Eve, which results in an observed error that requires Alice and Bob to perform error correction and privacy amplification. Both processes affect the secret key rate. The resulting key rate is given as \cite{PhysRevA.82.030301, PhysRevA.85.052310}
\begin{eqnarray}\label{12}
r_{\textrm{min}} & = &\log_2 d+\frac{d+1}{d} Q\log_2\left(\frac{Q}{d(d-1)}\right)\nonumber \\ & & +\left(1-\frac{d+1}{d}Q\right)\log_2\left(1-\frac{d+1}{d}Q\right),
\end{eqnarray}

where $Q$ is the average error rate from Eq. (\ref{9}). 

The secret key rate is given as the difference between the classical mutual information shared by Alice and Bob and the information shared by Alice and Eve as measured by the quantum mutual information. The quantum mutual information is also referred to as the Holevo quantity \cite{RevModPhys.81.1301}. The Holevo quantity measures the information that one has on Alice's data as a result of Eve's interaction with the signals as they pass to Bob. The secret key rate can be written as
\begin{equation}
r=I(A:B)-\chi(X:E),
\end{equation} 
where $I(A:B)$ is the classical mutual information and $\chi(X:E)=H(X)-S(E)-S(X,E)$ is the quantum mutual information or Holevo quantity where $H$ and $S$ denote the Shannon entropy and von Neumann entropy respectively. 
%The secure key rate is given in terms of the mutual information between Alice and Bob (i.e. the correlations between Alice and Bob's measurement results) and the mutual information between Alice and Eve (i.e. the measurement correlations between Alice and Eve) and it is the difference between these two mutual informations (i.e. the measurement correlations between Alice and Bob with the noise introduced by Eve removed).

The limit on the tolerable error rate that is safe for secret key generation can be improved by implementing a full set of ($d+1$) MUBs \cite{PhysRevA.85.052310, PhysRevLett.88.127902}. Using a full set of MUBs results in an increase in the tolerable error rate in which we can still extract a reasonable secret key without compromising the security of the protocol. However, this happens at the cost of reducing the transmission rate which is proportional to the probability $1/(d+1)$ that Alice and Bob choose the same basis. But in our protocol, this is not a problem since we make use of the asymmetric \cite{Lo2005} basis choice, so one does not pay the high cost of sifting with MUBs. 
%The limit on the average error rate that is safe for secret key generation can be improved by implementing a full set of ($d+1$) MUBs \cite{PhysRevA.85.052310, PhysRevLett.88.127902}}. Although the use of the full set of MUBs reduces the maximum limit on the average error rate in higher-dimensional QKD protocols, it is at the cost of reducing the transmission rate, which is proportional to the probability $1/(d+1)$ that Alice and Bob choose the same basis. 
In order to calculate the maximum tolerable error rate, $Q_{\textrm{max}}$, the secret key rate, $r_{\textrm{min}} $, is set to zero. 
%, meaning that below this value of $Q_{\textrm{max}} $ the protocol can still distribute a secure secret key (i.e. $r_{\textrm{min}}>0$). 
%As $Q$ tends to $Q_{\textrm{max}} $, more and more information is lost which results in Eve requiring less information ($r_{\textrm{min}}$) to break the security. 
%An analogue for this can be viewed as water leaking out of a pipe. If the leak (denoted by the error rate $Q$) is big, but we want the water loss to be low, then the flow rate of the water (denoted by the secret key rate $r_{\textrm{min}}$) must be reduced. Therefore if we wish to reduce the amount of information that Eve attains, however our channel has a high loss rate (or high average error rate $Q$), we need to decrease the secret key rate $r_{\textrm{min}}$. 

\section{Experimental Setup}
\begin{figure}[htbp]
\centerline{\includegraphics[scale=1.0]{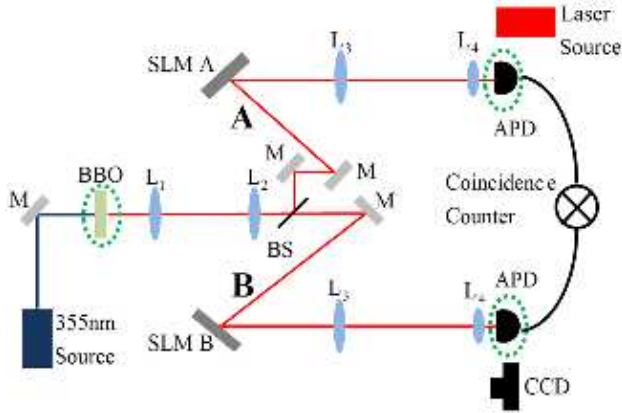}}
\caption{The experimental setup used to perform both the EB and P\&M QKD protocols. The plane of the crystal was relayed imaged onto SLMs A and B with the use of lenses, L$_{1}$ and L$_{2}$ (f$_{1}$ = 200 mm and f$_{2}$ = 400 mm). Lenses L$_{3}$ and L$_{4}$ (f$_{3}$ = 500 mm and f$_{4}$ = 2 mm) were used to relay image the SLM planes to single-mode fibres.}
\label{fig:setup}
\end{figure}

Our EB QKD protocol was implemented at the single photon level on entangled photon pairs depicted in Fig.~\ref{fig:setup}. A collimated 350 mW UV laser (Vanguard 355-2500) was directed to pump a 3 mm-thick type-I BBO crystal, producing collinear frequency-degenerate entangled photon pairs at 710 nm. A beam-splitter was used to separate the collinear signal and idler photons (depicted by arms A and B) which were directed and imaged (2$\times$) from the plane of the crystal onto spatial light modulators (SLMs) by a 4-\textit{f} telescope. The SLMs were used to execute the filter measurements and were encoded to manipulate both the phase and amplitude of the incident light \cite{Appl.Opt.47.2008, J.Opt.Soc.Am.A.24.2007, Appl.Opt.38.1999, Lima2011}, allowing only one particular superposition of the LG modes to be detected by the detector, while all the others are blocked.  False colour images of the types of filters (or holograms) encoded on the SLMs are presented in Fig.~\ref{fig:example}. The projected mode obtained at the plane of the SLM, be it either Gaussian or non-Gaussian, depending on whether the filter either does or does not match the state of the incident photon, was imaged (0.004$\times$) by a 4-\textit{f} telescope onto a single-mode fibre.  The fibres were connected to avalanche photodiodes which detected the photon pairs via a coincidence counter. The single count rates, $S_{A}$ and $S_{B}$, and the coincidence count rates, \textit{C}, were recorded simultaneously and accumulated over an integration time of 10 s.

An initial step in conducting our EB QKD protocol, was to test it classically in a P\&M based scheme.  Our experimental setup for the P\&M scheme can be illustrated with the use of Fig.~\ref{fig:setup} where the BBO crystal is considered to be reflective and the APD in Arm A is replaced with a laser source and the APD in Arm B with a CCD camera. This procedure is commonly referred to as back-projection or retrodiction \cite{Klyshko1988}. Conducting the protocol in this manner, provided a quicker and simpler method for the verification of the experimental procedure. 

\section{Results and Discussion}

\begin{figure}
\centerline{\includegraphics[scale=1.0]{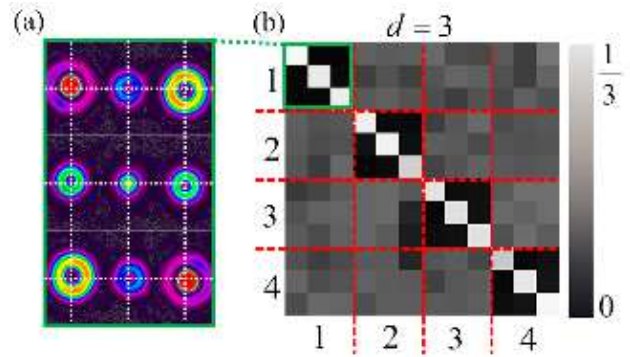}}
\caption{(a) Cross-sectional intensity profiles of the field recorded on the CCD for permutations of the first basis's states encoded on SLM A and SLM B. White cross-hairs mark the axis of propagation. (b) The normalized intensity recorded at the CCD when SLM A (Alice) and SLM B (Bob) select one of the three states from one of the 4 bases.}
\label{fig:d=3}
\end{figure}

By way of example we consider $d=3$ in our P\&M based protocol. We scanned through all possible states, defined by Eqs (\ref{single}) and (\ref{matrices}) and depicted in Fig.~\ref{fig:example}, on SLM A and SLM B. Figure~\ref{fig:d=3} (a) contains the cross-sectional intensity profiles recorded on the CCD (depicted in Fig.~\ref{fig:setup}) when SLM A and SLM B scanned through the states pertaining to the first basis. It is evident that when SLM A and SLM B select the same (different) states, a Gaussian mode (singularity) appears on axis. The normalized on-axis intensities are depicted in Fig.~\ref{fig:d=3} (b) for the permutation of all the bases elements for $d=3$. We note that the diagonal elements are equal to $1/3$ ($1/d$) and the elements corresponding to different bases are found to be $1/9$ ($1/d^2$). This validates the implementation of the filters (holograms) and their normalization. Our approach in obtaining the normalized joint probabilities is outlined in the Appendix.

\begin{figure}
\centerline{\includegraphics[scale=1.0]{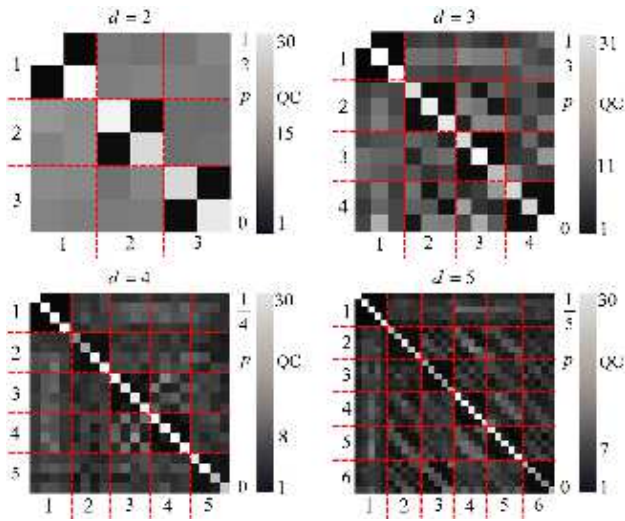}}
\caption{The normalized joint probabilities when SLM A (Alice) and SLM B (Bob) select one of the $d$ states from one of the $d+1$ bases for the EB scheme.}
\label{fig:All_ds}
\end{figure}

Following the successful implementation of the P\&M scheme, we proceeded to the EB scheme. For each permutation of the projective measurements by Alice and Bob in the EB scheme, the single count rates and coincidence count rates were recorded and the normalized joint probabilities calculated for $d=2,3, 4$ and 5 are given in Fig.~\ref{fig:All_ds}. In studying the data in Fig.~\ref{fig:All_ds}, it is evident that when the filter settings are the same, anti-correlations in all the bases are observed (denoted by the white diagonal elements).  In performing the projective measurements, completely orthogonal filter settings result in no correlations (an inconclusive measurement), while the overlap between the remaining filter settings is given as the inverse of the dimension (i.e. 1/d). 
%The intensities given in Fig.~\ref{fig:d=3} (b) and the coincidence count rates given in Fig.~\ref{fig:All_ds} are normalized such that the sum of the joint coincidence rates for measurements in any of the bases states is equivalent to unity. Thus resulting in the diagonal elements being equivalent to $1/d$  and the elements corresponding to measurements where $a \neq b$ are expected to be $1/d^2$. Our approach in obtaining the normalized joint probabilities is outlined in the Appendix.

\begin{figure}[here]
\centerline{\includegraphics[scale=1.0]{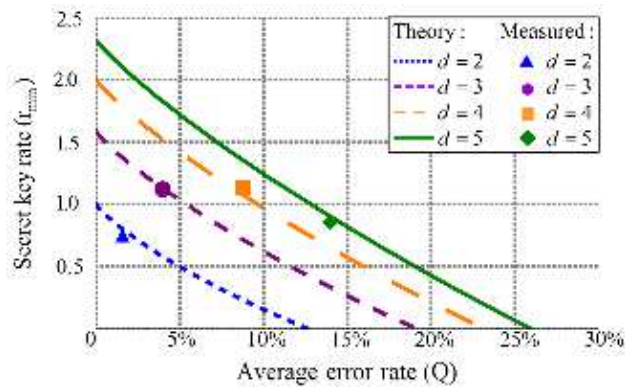}}
\caption{The secret key rate, $r_{\textrm{min}}$, as a function of the average error rate, $Q$, for different dimensions. The solid data points denote the measured values and the dashed curves the theoretical values calculated from Eq.~(\ref{12}).}
\label{fig:graph1}
\end{figure}

Based on the results from the normalized joint probabilities, we calculated the average error rate $Q$ according to Eq. (\ref{9}). We find that for $d$ = 2, 3, 4 and 5, the average error rate, $Q$ = 0.016, 0.040, 0.088, and 0.14, respectively. By using these values of $Q$ together with Eq.~(\ref{12}) we calculate the secret key rate to be $r_{\textrm{min}}$ = 0.7590, 1.123, 1.139 and 0.8606  for $d$ = 2, 3, 4 and 5, respectively. Figure~\ref{fig:graph1} contains the measured secret key rates plotted as a function of the measured average error rates for dimensions $d$ = 2, 3, 4 and 5, denoted by the data points. The curves denote the theoretical secret key rate as a function of the average error rate, plotted with the use of Eq.~(\ref{12}). For each dimension, $d$, the intersection between the dashed curves and the horizontal axis (i.e. where $r_{\textrm{min}}=0$) corresponds to the maximum permissible error rate ($Q_{\textrm{max}}$) in order to enable the secure distribution of a secret key. 
%It is at this point (where $r_{\textrm{min}}=0$) that Eve attains Bob's information (i.e. $I_{AB}=I_{AE}$). 
Ideally, we want to minimize the error rate $Q$ in order to maximize the secret key rate $r_{\textrm{min}}$. %Even though this is not the case,  all of our 4 measured average error rates do not exceed their maximum permissible error rates (illustrated in Fig.~\ref{fig:graph3} ), which confirms that the information can be sent securely between Alice and Bob.

\begin{figure}[here]
\centerline{\includegraphics[scale=1.0]{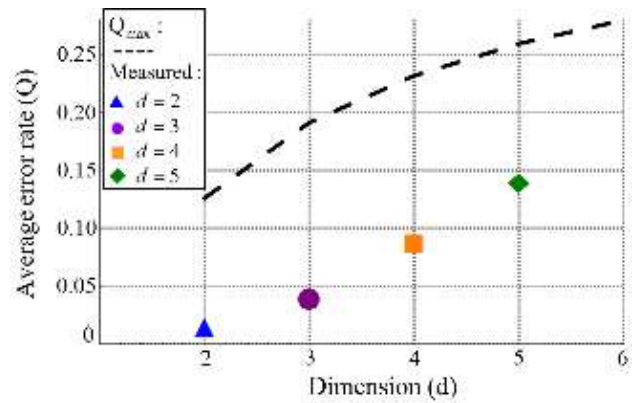}}
\caption{The measured average error rate ($Q$) and the maximum permissible error rate ($Q_{\textrm{max}}$) evaluated when $r_{\textrm{min}}=0$.}
\label{fig:graph3}
\end{figure}

\begin{figure}[here]
\centerline{\includegraphics[scale=1.0]{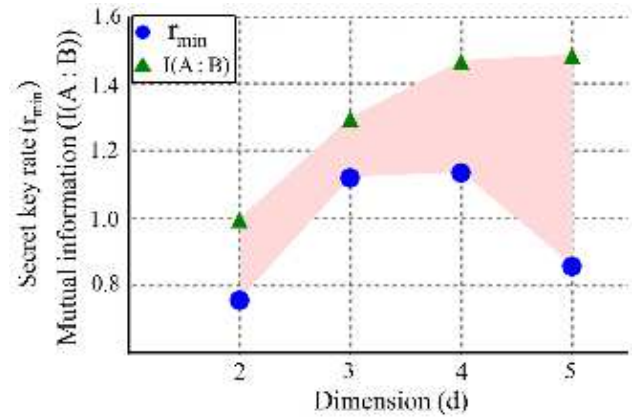}}
\caption{The Shannon mutual information I(A:B) \textit{(green)} and the secret key rate $r_{\textrm{min}}$ \textit{(blue)} plotted as a function of the dimension.}
\label{fig:graph2}
\end{figure}

The Shannon information for $d$ = 2, 3, 4 and 5 is calculated to be I(A:B) = 0.9999, 1.313, 1.478 and 1.487, respectively (depicted by the green data points in Fig.~\ref{fig:graph2}), while the Shannon mutual information increases monotonically, it seems to level off for $d$ = 4 and 5. On the other hand $r_{\textrm{min}}$ first increases and then decreases for $d$ = 5. This means that we have reached a finite limit on the dimension in which the protocol can encode, while still resulting in higher generation rates per photon. The difference between these two quantities (I(A:B) and $r_{\textrm{min}}$) is the mutual information between Alice and Eve, in other words the information that is shared between Alice and Eve (denoted by the red lines in Fig.~\ref{fig:graph2}). From our results it is evident that the noise (attributed to a disturbance by Eve) grows faster than the correlations between Alice and Bob that can be used to generate a key. As this is not expected theoretically, this may be due to the complexity associated with encoding higher-dimensional states holographically on pixelated, finite resolution, spatial light modulators. Our detection efficiency is low because our filter measurements are based on intensity masking and serve as a proof-of-principle experiment.
%Since there is no noise in the transmission channel, there is most probably a problem with the preparation or  measurement of our higher-dimensional states - an area for future investigations.  
  
\section{Conclusion}
%In this work, we have realized a P\&M and an entanglement-based QKD protocol for MUB measurements encoded in the OAM degree of freedom. We observe that encoding in higher-dimensional MUBs, leads to an increase in the encoding density per photon and increased key generation rates per photon. Furthermore, we have shown in the Appendix for the example of dimension $d=2$ that the detection efficiency depends on each basis and not on individual states within a basis, we assume that is true for all dimensions. This allows us to implement a standard MUB protocol with this approach.

In this work, we have realized a P\&M and an EB QKD protocol for $d = 2$ to 5 MUB measurements encoded in the OAM degree of freedom. We show that our protocol which is based on filter measurements can be mapped back into the original MUB protocol which uses full measurements.  In particular, we verify our claim that detection efficiency depends on a basis choice and not on the element within a basis, an important consideration for the protocol to work. We show this explicitly for $d=2$ and attest to the fact that this dependency holds for all dimensions. We infer from our measurements the average error rate, mutual information and secret key generation rate per photon for each dimension. We observe that encoding in higher-dimensional MUBs, leads to an increase in the encoding density per photon and increased key generation rates per photon.
This work is based upon research supported by the South African Research Chair Initiative of the Department of Science and Technology and National Research Foundation. T. Konrad. acknowledges the partial support from the National Research Foundation of South Africa [Grant No. 86325 (UID)]. M. Mafu especially thanks N. L\"{u}tkenhaus for hospitality and financial support during his stay at the Institute for Quantum Computing, University of Waterloo where part of the work was done and also thanks support from the South African Research Chair Initiative of the Department of Science and Technology and National Research Foundation.

% If in two-column mode, this environment will change to single-column
% format so that long equations can be displayed. Use
%\begin{widetext}
%\end{widetext}

% figures should be put into the text as floats.
% Use the graphics or graphicx packages (distributed with LaTeX2e)
% and the \includegraphics macro defined in those packages.
% See the LaTeX Graphics Companion by Michel Goosens, Sebastian Rahtz,
% entire page. There is no need to do explicit centering.

% \begin{figure}
% \end{figure}

%\appendix
%\section{}

%\begin{acknowledgments}
%\end{acknowledgments}


%\section*{References}
%\bibliographystyle{iopart-num}
\bibliography{iopart-num}

\begin{thebibliography}{32}%
\makeatletter
\providecommand \@ifxundefined [1]{%
 \@ifx{#1\undefined}
}%
\providecommand \@ifnum [1]{%
 \ifnum #1\expandafter \@firstoftwo
 \else \expandafter \@secondoftwo
 \fi
}%
\providecommand \@ifx [1]{%
 \ifx #1\expandafter \@firstoftwo
 \else \expandafter \@secondoftwo
 \fi
}%
\providecommand \natexlab [1]{#1}%
\providecommand \enquote  [1]{``#1''}%
\providecommand \bibnamefont  [1]{#1}%
\providecommand \bibfnamefont [1]{#1}%
\providecommand \citenamefont [1]{#1}%
\providecommand \href@noop [0]{\@secondoftwo}%
\providecommand \href [0]{\begingroup \@sanitize@url \@href}%
\providecommand \@href[1]{\@@startlink{#1}\@@href}%
\providecommand \@@href[1]{\endgroup#1\@@endlink}%
\providecommand \@sanitize@url [0]{\catcode `\\12\catcode `\$12\catcode
  `\&12\catcode `\#12\catcode `\^12\catcode `\_12\catcode `\%12\relax}%
\providecommand \@@startlink[1]{}%
\providecommand \@@endlink[0]{}%
\providecommand \url  [0]{\begingroup\@sanitize@url \@url }%
\providecommand \@url [1]{\endgroup\@href {#1}{\urlprefix }}%
\providecommand \urlprefix  [0]{URL }%
\providecommand \Eprint [0]{\href }%
\providecommand \doibase [0]{http://dx.doi.org/}%
\providecommand \selectlanguage [0]{\@gobble}%
\providecommand \bibinfo  [0]{\@secondoftwo}%
\providecommand \bibfield  [0]{\@secondoftwo}%
\providecommand \translation [1]{[#1]}%
\providecommand \BibitemOpen [0]{}%
\providecommand \bibitemStop [0]{}%
\providecommand \bibitemNoStop [0]{.\EOS\space}%
\providecommand \EOS [0]{\spacefactor3000\relax}%
\providecommand \BibitemShut  [1]{\csname bibitem#1\endcsname}%
\let\auto@bib@innerbib\@empty
%</preamble>
\bibitem [{\citenamefont {Gisin}\ \emph {et~al.}(2002)\citenamefont {Gisin},
  \citenamefont {Ribordy}, \citenamefont {Tittel},\ and\ \citenamefont
  {Zbinden}}]{RevModPhys.74.145}%
  \BibitemOpen
  \bibfield  {author} {\bibinfo {author} {\bibfnamefont {N.}~\bibnamefont
  {Gisin}}, \bibinfo {author} {\bibfnamefont {G.}~\bibnamefont {Ribordy}},
  \bibinfo {author} {\bibfnamefont {W.}~\bibnamefont {Tittel}}, \ and\ \bibinfo
  {author} {\bibfnamefont {H.}~\bibnamefont {Zbinden}},\ }\href {\doibase
  10.1103/RevModPhys.74.145} {\bibfield  {journal} {\bibinfo  {journal} {Rev.
  Mod. Phys.}\ }\textbf {\bibinfo {volume} {74}},\ \bibinfo {pages} {145}
  (\bibinfo {year} {2002})}\BibitemShut {NoStop}%
\bibitem [{\citenamefont {Scarani}\ \emph {et~al.}(2009)\citenamefont
  {Scarani}, \citenamefont {Bechmann-Pasquinucci}, \citenamefont {Cerf},
  \citenamefont {Du\ifmmode~\check{s}\else \v{s}\fi{}ek}, \citenamefont
  {L\"{u}tkenhaus},\ and\ \citenamefont {Peev}}]{RevModPhys.81.1301}%
  \BibitemOpen
  \bibfield  {author} {\bibinfo {author} {\bibfnamefont {V.}~\bibnamefont
  {Scarani}}, \bibinfo {author} {\bibfnamefont {H.}~\bibnamefont
  {Bechmann-Pasquinucci}}, \bibinfo {author} {\bibfnamefont {N.~J.}\
  \bibnamefont {Cerf}}, \bibinfo {author} {\bibfnamefont {M.}~\bibnamefont
  {Du\ifmmode~\check{s}\else \v{s}\fi{}ek}}, \bibinfo {author} {\bibfnamefont
  {N.}~\bibnamefont {L\"{u}tkenhaus}}, \ and\ \bibinfo {author} {\bibfnamefont
  {M.}~\bibnamefont {Peev}},\ }\href {\doibase 10.1103/RevModPhys.81.1301}
  {\bibfield  {journal} {\bibinfo  {journal} {Rev. Mod. Phys.}\ }\textbf
  {\bibinfo {volume} {81}},\ \bibinfo {pages} {1301} (\bibinfo {year}
  {2009})}\BibitemShut {NoStop}%
\bibitem [{\citenamefont {Bennett}\ \emph {et~al.}(1984)\citenamefont
  {Bennett}, \citenamefont {Brassard} \emph {et~al.}}]{C.H.Bennet}%
  \BibitemOpen
  \bibfield  {author} {\bibinfo {author} {\bibfnamefont {C.}~\bibnamefont
  {Bennett}}, \bibinfo {author} {\bibfnamefont {G.}~\bibnamefont {Brassard}},
  \emph {et~al.},\ }in\ \href@noop {} {\emph {\bibinfo {booktitle} {Proceedings
  of IEEE International Conference on Computers, Systems and Signal
  Processing}}},\ Vol.\ \bibinfo {volume} {175}\ (\bibinfo {organization}
  {Bangalore, India},\ \bibinfo {year} {1984})\BibitemShut {NoStop}%
\bibitem [{\citenamefont {Bennett}(1992)}]{PhysRevLett.68.3121}%
  \BibitemOpen
  \bibfield  {author} {\bibinfo {author} {\bibfnamefont {C.~H.}\ \bibnamefont
  {Bennett}},\ }\href {\doibase 10.1103/PhysRevLett.68.3121} {\bibfield
  {journal} {\bibinfo  {journal} {Phys. Rev. Lett.}\ }\textbf {\bibinfo
  {volume} {68}},\ \bibinfo {pages} {3121} (\bibinfo {year}
  {1992})}\BibitemShut {NoStop}%
\bibitem [{\citenamefont {Bru\ss{}}(1998)}]{PhysRevLett.81.3018}%
  \BibitemOpen
  \bibfield  {author} {\bibinfo {author} {\bibfnamefont {D.}~\bibnamefont
  {Bru\ss{}}},\ }\href {\doibase 10.1103/PhysRevLett.81.3018} {\bibfield
  {journal} {\bibinfo  {journal} {Phys. Rev. Lett.}\ }\textbf {\bibinfo
  {volume} {81}},\ \bibinfo {pages} {3018} (\bibinfo {year}
  {1998})}\BibitemShut {NoStop}%
\bibitem [{\citenamefont {Scarani}\ \emph {et~al.}(2004)\citenamefont
  {Scarani}, \citenamefont {Ac\'\i{}n}, \citenamefont {Ribordy},\ and\
  \citenamefont {Gisin}}]{PhysRevLett.92.057901}%
  \BibitemOpen
  \bibfield  {author} {\bibinfo {author} {\bibfnamefont {V.}~\bibnamefont
  {Scarani}}, \bibinfo {author} {\bibfnamefont {A.}~\bibnamefont {Ac\'\i{}n}},
  \bibinfo {author} {\bibfnamefont {G.}~\bibnamefont {Ribordy}}, \ and\
  \bibinfo {author} {\bibfnamefont {N.}~\bibnamefont {Gisin}},\ }\href
  {\doibase 10.1103/PhysRevLett.92.057901} {\bibfield  {journal} {\bibinfo
  {journal} {Phys. Rev. Lett.}\ }\textbf {\bibinfo {volume} {92}},\ \bibinfo
  {pages} {057901} (\bibinfo {year} {2004})}\BibitemShut {NoStop}%
\bibitem [{\citenamefont {Ekert}(1991)}]{ekert1991quantum}%
  \BibitemOpen
  \bibfield  {author} {\bibinfo {author} {\bibfnamefont {A.}~\bibnamefont
  {Ekert}},\ }\href@noop {} {\bibfield  {journal} {\bibinfo  {journal} {Phys.
  Rev. Lett.}\ }\textbf {\bibinfo {volume} {67}},\ \bibinfo {pages} {661}
  (\bibinfo {year} {1991})}\BibitemShut {NoStop}%
\bibitem [{\citenamefont {Schwinger}(1960)}]{schwinger1960unitary}%
  \BibitemOpen
  \bibfield  {author} {\bibinfo {author} {\bibfnamefont {J.}~\bibnamefont
  {Schwinger}},\ }\href@noop {} {\bibfield  {journal} {\bibinfo  {journal}
  {Proceedings of the national academy of sciences of the United States Of
  America}\ }\textbf {\bibinfo {volume} {46}},\ \bibinfo {pages} {570}
  (\bibinfo {year} {1960})}\BibitemShut {NoStop}%
\bibitem [{\citenamefont {Ivonovic}(1981)}]{ivonovic1981geometrical}%
  \BibitemOpen
  \bibfield  {author} {\bibinfo {author} {\bibfnamefont {I.}~\bibnamefont
  {Ivonovic}},\ }\href@noop {} {\bibfield  {journal} {\bibinfo  {journal}
  {Journal of Physics A: Mathematical and General}\ }\textbf {\bibinfo {volume}
  {14}},\ \bibinfo {pages} {3241} (\bibinfo {year} {1981})}\BibitemShut
  {NoStop}%
\bibitem [{\citenamefont {Wootters}\ and\ \citenamefont
  {Fields}(1989)}]{wootters1989optimal}%
  \BibitemOpen
  \bibfield  {author} {\bibinfo {author} {\bibfnamefont {W.}~\bibnamefont
  {Wootters}}\ and\ \bibinfo {author} {\bibfnamefont {B.}~\bibnamefont
  {Fields}},\ }\href@noop {} {\bibfield  {journal} {\bibinfo  {journal}
  {Annuals of Physics}\ }\textbf {\bibinfo {volume} {191}},\ \bibinfo {pages}
  {363} (\bibinfo {year} {1989})}\BibitemShut {NoStop}%
\bibitem [{\citenamefont {Adamson}\ and\ \citenamefont
  {Steinberg}(2010)}]{PhysRevLett.105.030406}%
  \BibitemOpen
  \bibfield  {author} {\bibinfo {author} {\bibfnamefont {R.~B.~A.}\
  \bibnamefont {Adamson}}\ and\ \bibinfo {author} {\bibfnamefont {A.~M.}\
  \bibnamefont {Steinberg}},\ }\href {\doibase 10.1103/PhysRevLett.105.030406}
  {\bibfield  {journal} {\bibinfo  {journal} {Phys. Rev. Lett.}\ }\textbf
  {\bibinfo {volume} {105}},\ \bibinfo {pages} {030406} (\bibinfo {year}
  {2010})}\BibitemShut {NoStop}%
\bibitem [{\citenamefont {Fern\'andez-P\'erez}\ \emph
  {et~al.}(2011)\citenamefont {Fern\'andez-P\'erez}, \citenamefont {Klimov},\
  and\ \citenamefont {Saavedra}}]{PhysRevA.83.052332}%
  \BibitemOpen
  \bibfield  {author} {\bibinfo {author} {\bibfnamefont {A.}~\bibnamefont
  {Fern\'andez-P\'erez}}, \bibinfo {author} {\bibfnamefont {A.~B.}\
  \bibnamefont {Klimov}}, \ and\ \bibinfo {author} {\bibfnamefont
  {C.}~\bibnamefont {Saavedra}},\ }\href {\doibase 10.1103/PhysRevA.83.052332}
  {\bibfield  {journal} {\bibinfo  {journal} {Phys. Rev. A}\ }\textbf {\bibinfo
  {volume} {83}},\ \bibinfo {pages} {052332} (\bibinfo {year}
  {2011})}\BibitemShut {NoStop}%
\bibitem [{\citenamefont {Giovannini}\ \emph {et~al.}(2013)\citenamefont
  {Giovannini}, \citenamefont {Romero}, \citenamefont {Leach}, \citenamefont
  {Dudley}, \citenamefont {Forbes},\ and\ \citenamefont
  {Padgett}}]{giovanni2013}%
  \BibitemOpen
  \bibfield  {author} {\bibinfo {author} {\bibfnamefont {D.}~\bibnamefont
  {Giovannini}}, \bibinfo {author} {\bibfnamefont {J.}~\bibnamefont {Romero}},
  \bibinfo {author} {\bibfnamefont {J.}~\bibnamefont {Leach}}, \bibinfo
  {author} {\bibfnamefont {A.}~\bibnamefont {Dudley}}, \bibinfo {author}
  {\bibfnamefont {A.}~\bibnamefont {Forbes}}, \ and\ \bibinfo {author}
  {\bibfnamefont {M.~J.}\ \bibnamefont {Padgett}},\ }\href@noop {} {\bibfield
  {journal} {\bibinfo  {journal} {Phys. Rev. Lett.}\ }\textbf {\bibinfo
  {volume} {110}},\ \bibinfo {pages} {143601} (\bibinfo {year}
  {2013})}\BibitemShut {NoStop}%
\bibitem [{\citenamefont {Calderbank}\ \emph {et~al.}(1997)\citenamefont
  {Calderbank}, \citenamefont {Rains}, \citenamefont {Shor},\ and\
  \citenamefont {Sloane}}]{calderbank1997quantum}%
  \BibitemOpen
  \bibfield  {author} {\bibinfo {author} {\bibfnamefont {A.}~\bibnamefont
  {Calderbank}}, \bibinfo {author} {\bibfnamefont {E.}~\bibnamefont {Rains}},
  \bibinfo {author} {\bibfnamefont {P.}~\bibnamefont {Shor}}, \ and\ \bibinfo
  {author} {\bibfnamefont {N.}~\bibnamefont {Sloane}},\ }\href@noop {}
  {\bibfield  {journal} {\bibinfo  {journal} {Phys. Rev. Lett.}\ }\textbf
  {\bibinfo {volume} {78}},\ \bibinfo {pages} {405} (\bibinfo {year}
  {1997})}\BibitemShut {NoStop}%
\bibitem [{\citenamefont {Gottesman}(1996)}]{PhysRevA.54.1862}%
  \BibitemOpen
  \bibfield  {author} {\bibinfo {author} {\bibfnamefont {D.}~\bibnamefont
  {Gottesman}},\ }\href {\doibase 10.1103/PhysRevA.54.1862} {\bibfield
  {journal} {\bibinfo  {journal} {Phys. Rev. A}\ }\textbf {\bibinfo {volume}
  {54}},\ \bibinfo {pages} {1862} (\bibinfo {year} {1996})}\BibitemShut
  {NoStop}%
\bibitem [{\citenamefont {Barnett}(2009)}]{barnett2009}%
  \BibitemOpen
  \bibfield  {author} {\bibinfo {author} {\bibfnamefont {S.~M.}\ \bibnamefont
  {Barnett}},\ }\href@noop {} {\emph {\bibinfo {title} {Quantum Information}}}\
  (\bibinfo  {publisher} {Oxford University Press},\ \bibinfo {year}
  {2009})\BibitemShut {NoStop}%
\bibitem [{\citenamefont {Durt}\ \emph {et~al.}(2010)\citenamefont {Durt},
  \citenamefont {Englert}, \citenamefont {Bengtsson},\ and\ \citenamefont
  {Zyczkowski}}]{durt2010}%
  \BibitemOpen
  \bibfield  {author} {\bibinfo {author} {\bibfnamefont {T.}~\bibnamefont
  {Durt}}, \bibinfo {author} {\bibfnamefont {B.-G.}\ \bibnamefont {Englert}},
  \bibinfo {author} {\bibfnamefont {I.}~\bibnamefont {Bengtsson}}, \ and\
  \bibinfo {author} {\bibfnamefont {K.}~\bibnamefont {Zyczkowski}},\
  }\href@noop {} {\bibfield  {journal} {\bibinfo  {journal} {International
  Journal of Quantum Information}\ }\textbf {\bibinfo {volume} {08}},\ \bibinfo
  {pages} {535} (\bibinfo {year} {2010})}\BibitemShut {NoStop}%
\bibitem [{\citenamefont {Sheridan}\ and\ \citenamefont
  {Scarani}(2010)}]{PhysRevA.82.030301}%
  \BibitemOpen
  \bibfield  {author} {\bibinfo {author} {\bibfnamefont {L.}~\bibnamefont
  {Sheridan}}\ and\ \bibinfo {author} {\bibfnamefont {V.}~\bibnamefont
  {Scarani}},\ }\href {\doibase 10.1103/PhysRevA.82.030301} {\bibfield
  {journal} {\bibinfo  {journal} {Phys. Rev. A}\ }\textbf {\bibinfo {volume}
  {82}},\ \bibinfo {pages} {030301} (\bibinfo {year} {2010})}\BibitemShut
  {NoStop}%
\bibitem [{\citenamefont {Ferenczi}\ and\ \citenamefont
  {L\"utkenhaus}(2012)}]{PhysRevA.85.052310}%
  \BibitemOpen
  \bibfield  {author} {\bibinfo {author} {\bibfnamefont {A.}~\bibnamefont
  {Ferenczi}}\ and\ \bibinfo {author} {\bibfnamefont {N.}~\bibnamefont
  {L\"utkenhaus}},\ }\href {\doibase 10.1103/PhysRevA.85.052310} {\bibfield
  {journal} {\bibinfo  {journal} {Phys. Rev. A}\ }\textbf {\bibinfo {volume}
  {85}},\ \bibinfo {pages} {052310} (\bibinfo {year} {2012})}\BibitemShut
  {NoStop}%
\bibitem [{\citenamefont {Allen}\ \emph {et~al.}(1992)\citenamefont {Allen},
  \citenamefont {Beijersbergen}, \citenamefont {Spreeuw},\ and\ \citenamefont
  {Woerdman}}]{Allen1992}%
  \BibitemOpen
  \bibfield  {author} {\bibinfo {author} {\bibfnamefont {L.}~\bibnamefont
  {Allen}}, \bibinfo {author} {\bibfnamefont {M.}~\bibnamefont
  {Beijersbergen}}, \bibinfo {author} {\bibfnamefont {R.}~\bibnamefont
  {Spreeuw}}, \ and\ \bibinfo {author} {\bibfnamefont {J.}~\bibnamefont
  {Woerdman}},\ }\href@noop {} {\bibfield  {journal} {\bibinfo  {journal}
  {\pra}\ }\textbf {\bibinfo {volume} {45}},\ \bibinfo {pages} {8185} (\bibinfo
  {year} {1992})}\BibitemShut {NoStop}%
\bibitem [{\citenamefont {Bechmann-Pasquinucci}\ and\ \citenamefont
  {Peres}(2000)}]{bechmann2000quantum}%
  \BibitemOpen
  \bibfield  {author} {\bibinfo {author} {\bibfnamefont {H.}~\bibnamefont
  {Bechmann-Pasquinucci}}\ and\ \bibinfo {author} {\bibfnamefont
  {A.}~\bibnamefont {Peres}},\ }\href@noop {} {\bibfield  {journal} {\bibinfo
  {journal} {Phys. Rev. Lett.}\ }\textbf {\bibinfo {volume} {85}},\ \bibinfo
  {pages} {3313} (\bibinfo {year} {2000})}\BibitemShut {NoStop}%
\bibitem [{\citenamefont {Gr{\"o}blacher}\ \emph {et~al.}(2006)\citenamefont
  {Gr{\"o}blacher}, \citenamefont {Jennewein}, \citenamefont {Vaziri},
  \citenamefont {Weihs},\ and\ \citenamefont
  {Zeilinger}}]{groblacher2006experimental}%
  \BibitemOpen
  \bibfield  {author} {\bibinfo {author} {\bibfnamefont {S.}~\bibnamefont
  {Gr{\"o}blacher}}, \bibinfo {author} {\bibfnamefont {T.}~\bibnamefont
  {Jennewein}}, \bibinfo {author} {\bibfnamefont {A.}~\bibnamefont {Vaziri}},
  \bibinfo {author} {\bibfnamefont {G.}~\bibnamefont {Weihs}}, \ and\ \bibinfo
  {author} {\bibfnamefont {A.}~\bibnamefont {Zeilinger}},\ }\href@noop {}
  {\bibfield  {journal} {\bibinfo  {journal} {New Journal of Physics}\ }\textbf
  {\bibinfo {volume} {8}},\ \bibinfo {pages} {75} (\bibinfo {year}
  {2006})}\BibitemShut {NoStop}%
\bibitem [{\citenamefont {Yu}\ \emph {et~al.}(2008)\citenamefont {Yu},
  \citenamefont {Lin},\ and\ \citenamefont {Huang}}]{PhysRevA.78.012344}%
  \BibitemOpen
  \bibfield  {author} {\bibinfo {author} {\bibfnamefont {I.-C.}\ \bibnamefont
  {Yu}}, \bibinfo {author} {\bibfnamefont {F.-L.}\ \bibnamefont {Lin}}, \ and\
  \bibinfo {author} {\bibfnamefont {C.-Y.}\ \bibnamefont {Huang}},\ }\href
  {\doibase 10.1103/PhysRevA.78.012344} {\bibfield  {journal} {\bibinfo
  {journal} {Phys. Rev. A}\ }\textbf {\bibinfo {volume} {78}},\ \bibinfo
  {pages} {012344} (\bibinfo {year} {2008})}\BibitemShut {NoStop}%
\bibitem [{\citenamefont {Cerf}\ \emph {et~al.}(2002)\citenamefont {Cerf},
  \citenamefont {Bourennane}, \citenamefont {Karlsson},\ and\ \citenamefont
  {Gisin}}]{PhysRevLett.88.127902}%
  \BibitemOpen
  \bibfield  {author} {\bibinfo {author} {\bibfnamefont {N.~J.}\ \bibnamefont
  {Cerf}}, \bibinfo {author} {\bibfnamefont {M.}~\bibnamefont {Bourennane}},
  \bibinfo {author} {\bibfnamefont {A.}~\bibnamefont {Karlsson}}, \ and\
  \bibinfo {author} {\bibfnamefont {N.}~\bibnamefont {Gisin}},\ }\href
  {\doibase 10.1103/PhysRevLett.88.127902} {\bibfield  {journal} {\bibinfo
  {journal} {Phys. Rev. Lett.}\ }\textbf {\bibinfo {volume} {88}},\ \bibinfo
  {pages} {127902} (\bibinfo {year} {2002})}\BibitemShut {NoStop}%
\bibitem [{\citenamefont {Rodenburg}\ \emph {et~al.}(2012)\citenamefont
  {Rodenburg}, \citenamefont {Lavery}, \citenamefont {Malik}, \citenamefont
  {O'Sullivan}, \citenamefont {Mirhosseini}, \citenamefont {Robertson},
  \citenamefont {Padgett},\ and\ \citenamefont {Boyd}}]{BoydQKD}%
  \BibitemOpen
  \bibfield  {author} {\bibinfo {author} {\bibfnamefont {B.}~\bibnamefont
  {Rodenburg}}, \bibinfo {author} {\bibfnamefont {M.~J.~P.}\ \bibnamefont
  {Lavery}}, \bibinfo {author} {\bibfnamefont {M.}~\bibnamefont {Malik}},
  \bibinfo {author} {\bibfnamefont {M.~N.}\ \bibnamefont {O'Sullivan}},
  \bibinfo {author} {\bibfnamefont {M.}~\bibnamefont {Mirhosseini}}, \bibinfo
  {author} {\bibfnamefont {D.~J.}\ \bibnamefont {Robertson}}, \bibinfo {author}
  {\bibfnamefont {M.~J.}\ \bibnamefont {Padgett}}, \ and\ \bibinfo {author}
  {\bibfnamefont {R.~W.}\ \bibnamefont {Boyd}},\ }\href@noop {} {\bibfield
  {journal} {\bibinfo  {journal} {Opt. Lett.}\ }\textbf {\bibinfo {volume}
  {37}},\ \bibinfo {pages} {3735} (\bibinfo {year} {2012})}\BibitemShut
  {NoStop}%
\bibitem [{\citenamefont {Lo}\ \emph {et~al.}(2005)\citenamefont {Lo},
  \citenamefont {Chau},\ and\ \citenamefont {Ardehali}}]{Lo2005}%
  \BibitemOpen
  \bibfield  {author} {\bibinfo {author} {\bibfnamefont {H.}~\bibnamefont
  {Lo}}, \bibinfo {author} {\bibfnamefont {H.}~\bibnamefont {Chau}}, \ and\
  \bibinfo {author} {\bibfnamefont {M.}~\bibnamefont {Ardehali}},\ }\href@noop
  {} {\bibfield  {journal} {\bibinfo  {journal} {Journal of Cryptology}\
  }\textbf {\bibinfo {volume} {18}},\ \bibinfo {pages} {133} (\bibinfo {year}
  {2005})}\BibitemShut {NoStop}%
\bibitem [{\citenamefont {Gruneisen}\ \emph {et~al.}(2008)\citenamefont
  {Gruneisen}, \citenamefont {Miller}, \citenamefont {Dymale},\ and\
  \citenamefont {Sweiti}}]{Appl.Opt.47.2008}%
  \BibitemOpen
  \bibfield  {author} {\bibinfo {author} {\bibfnamefont {M.~T.}\ \bibnamefont
  {Gruneisen}}, \bibinfo {author} {\bibfnamefont {W.~A.}\ \bibnamefont
  {Miller}}, \bibinfo {author} {\bibfnamefont {R.~C.}\ \bibnamefont {Dymale}},
  \ and\ \bibinfo {author} {\bibfnamefont {A.~M.}\ \bibnamefont {Sweiti}},\
  }\href@noop {} {\bibfield  {journal} {\bibinfo  {journal} {Applied Optics}\
  }\textbf {\bibinfo {volume} {47}},\ \bibinfo {pages} {A32} (\bibinfo {year}
  {2008})}\BibitemShut {NoStop}%
\bibitem [{\citenamefont {Arriz\'{o}n}\ \emph {et~al.}(2007)\citenamefont
  {Arriz\'{o}n}, \citenamefont {Ruiz}, \citenamefont {Carrada},\ and\
  \citenamefont {Gonz\'{a}lez}}]{J.Opt.Soc.Am.A.24.2007}%
  \BibitemOpen
  \bibfield  {author} {\bibinfo {author} {\bibfnamefont {V.}~\bibnamefont
  {Arriz\'{o}n}}, \bibinfo {author} {\bibfnamefont {U.}~\bibnamefont {Ruiz}},
  \bibinfo {author} {\bibfnamefont {R.}~\bibnamefont {Carrada}}, \ and\
  \bibinfo {author} {\bibfnamefont {A.}~\bibnamefont {Gonz\'{a}lez}},\
  }\href@noop {} {\bibfield  {journal} {\bibinfo  {journal} {J. Opt. Soc. Am.
  A}\ }\textbf {\bibinfo {volume} {24}},\ \bibinfo {pages} {3500} (\bibinfo
  {year} {2007})}\BibitemShut {NoStop}%
\bibitem [{\citenamefont {Davies}\ \emph {et~al.}(1999)\citenamefont {Davies},
  \citenamefont {Cottrell}, \citenamefont {Campos}, \citenamefont {Yzuel},\
  and\ \citenamefont {Moreno}}]{Appl.Opt.38.1999}%
  \BibitemOpen
  \bibfield  {author} {\bibinfo {author} {\bibfnamefont {J.~A.}\ \bibnamefont
  {Davies}}, \bibinfo {author} {\bibfnamefont {D.~M.}\ \bibnamefont
  {Cottrell}}, \bibinfo {author} {\bibfnamefont {J.}~\bibnamefont {Campos}},
  \bibinfo {author} {\bibfnamefont {M.~J.}\ \bibnamefont {Yzuel}}, \ and\
  \bibinfo {author} {\bibfnamefont {I.}~\bibnamefont {Moreno}},\ }\href@noop {}
  {\bibfield  {journal} {\bibinfo  {journal} {Applied Optics}\ }\textbf
  {\bibinfo {volume} {38}},\ \bibinfo {pages} {5004} (\bibinfo {year}
  {1999})}\BibitemShut {NoStop}%
\bibitem [{\citenamefont {Lima}\ \emph {et~al.}(2011)\citenamefont {Lima},
  \citenamefont {Neves}, \citenamefont {Guzm\'{a}n}, \citenamefont {G\'{o}mez},
  \citenamefont {Nogueira}, \citenamefont {Delgado}, \citenamefont {Vargas},\
  and\ \citenamefont {C.}}]{Lima2011}%
  \BibitemOpen
  \bibfield  {author} {\bibinfo {author} {\bibfnamefont {G.}~\bibnamefont
  {Lima}}, \bibinfo {author} {\bibfnamefont {L.}~\bibnamefont {Neves}},
  \bibinfo {author} {\bibfnamefont {R.}~\bibnamefont {Guzm\'{a}n}}, \bibinfo
  {author} {\bibfnamefont {E.~S.}\ \bibnamefont {G\'{o}mez}}, \bibinfo {author}
  {\bibfnamefont {W.~A.~T.}\ \bibnamefont {Nogueira}}, \bibinfo {author}
  {\bibfnamefont {A.}~\bibnamefont {Delgado}}, \bibinfo {author} {\bibfnamefont
  {A.}~\bibnamefont {Vargas}}, \ and\ \bibinfo {author} {\bibfnamefont
  {S.}~\bibnamefont {C.}},\ }\href@noop {} {\bibfield  {journal} {\bibinfo
  {journal} {Opt. Express}\ }\textbf {\bibinfo {volume} {19}},\ \bibinfo
  {pages} {3542} (\bibinfo {year} {2011})}\BibitemShut {NoStop}%
\bibitem [{\citenamefont {Klyshko}(1988)}]{Klyshko1988}%
  \BibitemOpen
  \bibfield  {author} {\bibinfo {author} {\bibfnamefont {D.}~\bibnamefont
  {Klyshko}},\ }\href@noop {} {\bibfield  {journal} {\bibinfo  {journal}
  {Soviet Physics Uspekhi}\ }\textbf {\bibinfo {volume} {31}},\ \bibinfo
  {pages} {74} (\bibinfo {year} {1988})}\BibitemShut {NoStop}%
\bibitem [{\citenamefont {Jennewein}\ \emph {et~al.}(2011)\citenamefont
  {Jennewein}, \citenamefont {Barbieri},\ and\ \citenamefont
  {White}}]{Jennewein2010}%
  \BibitemOpen
  \bibfield  {author} {\bibinfo {author} {\bibfnamefont {T.}~\bibnamefont
  {Jennewein}}, \bibinfo {author} {\bibfnamefont {M.}~\bibnamefont {Barbieri}},
  \ and\ \bibinfo {author} {\bibfnamefont {A.~G.}\ \bibnamefont {White}},\
  }\href@noop {} {\bibfield  {journal} {\bibinfo  {journal} {J. Mod. Phys.}\
  }\textbf {\bibinfo {volume} {58}},\ \bibinfo {pages} {276} (\bibinfo {year}
  {2011})}\BibitemShut {NoStop}%
\end{thebibliography}

\section{Appendix}

\subsection{Shannon information}
In order to analyze the security of our scheme, we employ the concept of mutual information given by Shannon. The Shannon entropy gives a measure of uncertainty for a random variable $A$ with alphabet $\mathcal{A}$ and is defined as $H(A)=-\sum_{a\in\mathcal{A}}{p(a)\log_2p(a)},$
where $p(a)$ is the probability of outcome $a$. 
The classical mutual information is defined as the amount by which the Shannon entropy on $A$ decreases when one learns about $B$. The classical mutual information $I(A:B)$, gives a degree of correlation between Alice (\textit{A}), and Bob's (\textit{B}) data and it is also an upper bound on the secret key rate. It is defined as $
I(A:B)=H(A)+H(B)-H(A,B),$ where $H(A,B)$ is the joint entropy. The joint entropy is used to measure the total uncertainty about the pair $(A,B)$. It is expressed as $
 H(A,B)=-\sum_{a\in \mathcal{A}}\sum_{b\in \mathcal{B}}p(a,b)\log_2 p(a,b).$
 After data-processing, Alice and Bob apply a key map where their respective data is mapped to raw keys $K$ and $K'$. In this step, the total probability distribution remains unchanged but the total classical mutual information changes to $I(A':B)$, which is expressed as
 \begin{equation}
 I(A':B)=H(A')+H(B)-H(A',B),
 \end{equation} 
 where $H(A')=\sum_a\sum_{ij}p_{ij}^{aa}\log_2 p_{ij}^{aa}$. The joint entropy is defined in a similar manner as above. 

\subsection{Calculation of detection efficiencies}

In this section, we show how to formalize and verify the claim that the detection efficiencies depend only on the bases but are the same for all elements within a basis. We demonstrate the calculation of detection efficiencies by comparing the expected and detected number of clicks for the case of qubit pairs ($d=2$). For this purpose, we first calculate the expected number of detection events by following the light beam from the laser source to the detection device. Afterwords we relate them to the measured counts. By comparing the single count rates and the coincidence count rates we obtain an expression for the detection efficiency for each basis state.  
   
\subsubsection{Photon pair creation and action of the beam splitter} 
The state of the light  exiting the laser source can be represented by a coherent state with complex parameter $\alpha$ which specifies the intensity and phase of the light:
\begin{align}
\ket{\alpha} &= \mathcal{D}(\alpha)\ket{0} =  \exp(\alpha b_0^\dagger - \alpha^* b_0) \ket{0},
\end{align}
where $\ket{0}$ is the vacuum state, $b_0$ and  $b_0^\dagger$ are annihilation and creation operators, respectively, with index  referring to OAM value $l=0$.    The operator $\mathcal{D}(\alpha)$ is called a displacement operator. The laser beam pumps a BBO crystal, creating pairs of photons with OAM values $\pm l$ by type I parametric down conversion. This process can be modeled by the following transformation of creation operators
\begin{align}
b_0^\dagger \to  \sum_\ell\sqrt{\chi_\ell} a_\ell^\dagger a_{-\ell}^\dagger,
\label{pdc}
\end{align}
where  $\chi_\ell$ is the creation probability of a  photon pair with OAM values $\pm\ell$ and $a^\dagger_{\pm\ell}$ are the corresponding creation operators. After passing through the BBO crystal the light is sent to a $50:50$ beam splitter resulting in the transformation:
\begin{align}
a^\dagger_\ell \to \frac{1}{\sqrt{2}}\left(a_{\ell, A}^\dagger + a_{\ell, B}^\dagger \right),
\end{align}
where $A$ and $B$  refer to the two beams exiting the beam splitter.
Thus, the combined action of the BBO crystal and the beam splitter reads
\begin{align}
a_0^\dagger &\to \sum_{\ell = 0}^\infty\sqrt{\chi_\ell}\left( \frac{a_{\ell,A}^\dagger + a_{\ell,B}^\dagger}{\sqrt{2}}\right)\left(\frac{a_{-\ell,A }^\dagger +a_{-\ell,B}^\dagger}{\sqrt{2}}\right).
\end{align}
It maps the displacement operator $\mathcal{D}(\alpha)$ to a squeeze operator $\mathcal{S}(\alpha\sqrt{\chi_\ell})$ given by
\begin{align}
\mathcal{S}(\alpha \sqrt{\chi_\ell}) &= \exp\left(\alpha\sum_{\ell=0}^\infty\sqrt{\chi_\ell}\left( \frac{a_{\ell,A}^\dagger + a_{\ell,B}^\dagger}{\sqrt{2}}\right) \right.\nonumber\\
& \left.\times \left(\frac{a_{-\ell,A }^\dagger +a_{-\ell,B}^\dagger}{\sqrt{2}}\right)- \alpha^* \sum_{\ell=0}^\infty\sqrt{\chi_\ell} \right.\nonumber\\
& \left.\times\left( \frac{a_{\ell,A} + a_{\ell,B}}{\sqrt{2}}\right)\left(\frac{a_{-\ell,A } + a_{-\ell,B}}{\sqrt{2}}\right)\right).
\end{align}
 Thus, the initial coherent state is transformed into a (two-mode) squeezed vacuum state:  $\ket{\tilde\alpha} = \mathcal{S}(\alpha\sqrt{\chi_\ell})\ket{0}$. For small value of $\alpha\sqrt{\chi_\ell}$ the state $\ket{\tilde\alpha}$ can be approximated to the first order in $\alpha\sqrt{\chi_\ell}$ as:
\begin{align}
\ket{\tilde\alpha} &\approx \mathcal{N} \left[1 + \alpha\sum_{\ell =0}^\infty \sqrt{\chi_\ell} \left(\frac{a_{\ell,A}^\dagger + a_{\ell,B}^\dagger}{\sqrt{2}}\right)\right.\nonumber\\
& \left.\times \left(\frac{a_{-\ell,A}^\dagger + a_{-\ell,B}^\dagger}{\sqrt{2}}\right)\right]\ket{0},
\end{align}
where $\mathcal{N}$ is the normalization constant. The vacuum does not play any role as far as photon detections are concerned, thus, one can ignore the vacuum component. This results in the (unnormalized) state $\ket{\psi}$ which reads:

\begin{align}
\ket{\psi} &= \alpha\sum_{\ell =0}^\infty \sqrt{\chi_\ell} \left(\frac{a_{\ell,A}^\dagger + a_{\ell,B}^\dagger}{\sqrt{2}}\right)\left(\frac{a_{-\ell,A}^\dagger + a_{-\ell,B}^\dagger}{\sqrt{2}}\right)\ket{0},\\
&= \frac{\alpha}{2}\sum_{\ell =0}^\infty \sqrt{\chi_\ell}(a_{\ell,A}^\dagger a_{-\ell,A}^\dagger + a_{\ell,A}^\dagger a_{-\ell,B}^\dagger\nonumber\\
&\qquad\qquad\qquad + a_{\ell,B}^\dagger a_{-\ell,A}^\dagger + a_{\ell,B}^\dagger a_{-\ell,B}^\dagger)\ket{0},\label{eqn-21}
\end{align}
%\SKG{Therefore, $\ket{\psi}$ is the combined state of the beams $A$ and $B$ which will be measured at the detectors $A$ and $B$ with different filter settings. }

\subsubsection{Measurements}
After the BBO crystal and the beam spitter filter measurements projecting onto individual basis modes were carried out independently in both beams  $A$ and  $B$. The signal for each basis mode was detected by means of avalanche photodiodes. These detectors  respond to incident photons, but do not discriminate between a single photon and multiple photons. However, the probability for a click varies for different photon numbers. The probability to obtain 
a click  in a filter measurement of mode $s$ can be modeled by the expectation value of the effect $P_s$ defined by
\begin{align}
P_s &= \sum_{n =1}^\infty \eta_s^{(n)}\ket{n_s}\bra{n_s},
\end{align}
where  $\eta^{(n)}_s$ represents the probability for  $n$ photons in mode $s$ to trigger a detector click and reads \cite{Jennewein2010}
\begin{align}
\eta^{(n)}_s &= 1 - (1-\eta^{(1)}_s)^n,\nonumber\\
&\approx n\eta_s^{(1)} ~\mbox{ for small  }~ \eta_s^{(1)}.
\end{align} 
Because of photon loss on the path from source to detector and non-ideal detection, only a fraction of the detection events expected under ideal conditions is measured in the experiment. We attribute any loss to non-ideal detection.
The probability of coincidence can be calculated as an expectation value of  the operator $P_s\otimes P_{s'}$ with respect to the state $\ket{\psi}$ (cp. Eq.\  \eqref{eqn-21}) after the beam splitter. 

From Eq.~\eqref{eqn-21} it is clear that  only the  single photon components of state $\ket{\psi}$ can yield a click of detector $A$ for OAM value $\ell$, leading to a detection probability  of \begin{align} p_{\ell,A} = \bra{\psi} P_\ell \otimes \mathbb{I} \ket{\psi} =  \eta^{(1)}_{\ell,A}|\alpha|^2\chi_\ell/2\,. \end{align}
 %Since the state stays invariant under permutation of  modes $A$ and  $B$, as well as  OAM values $\ell$ and  $-\ell$, the probability of a detection event in  $A$ or $B$ is the same for both OAM values $\ell$ and $-\ell$: $$p_{\ell,A} \,= p_{-\ell,A}\,=p_{\ell,B} \,= p_{-\ell,B}\,.$$
Similarly, we can calculate the other probabilities as:
\begin{align}
p_{-\ell,A} &= \bra{\psi} P_{-\ell} \otimes \mathbb{I} \ket{\psi} = \eta^{(1)}_{-\ell,A}|\alpha|^2\chi_\ell/2\,,\\
p_{\ell,B} &= \bra{\psi} \mathbb{I} \otimes P_\ell  \ket{\psi} = \eta^{(1)}_{\ell,B}|\alpha|^2\chi_\ell/2\,,\\
p_{-\ell,B} &= \bra{\psi} \mathbb{I} \otimes P_{-\ell}  \ket{\psi} = \eta^{(1)}_{-\ell,B}|\alpha|^2\chi_\ell/2\,.
\end{align}

The probability of the coincidence count in  detector $A$ with OAM  value $\ell$ and in detector $B$ with OAM value  $-\ell$ amounts to 
\begin{align}p_{\ell,A,-\ell,B} = \bra{\psi} P_{\ell} \otimes P_{-\ell} \ket{\psi} = \eta^{(1)}_{\ell,A}\eta^{(1)}_{-\ell,B}|\alpha|^2\chi_\ell/4\,. \end{align}

For the measured count of  clicks $C_{\ell,A} $ in detector $A$ with OAM value $\ell$, and the measured count $C_{-\ell,B}$ in detector $B$ with OAM value $-\ell$  we obtain the expressions:
\begin{align}
C_{\ell,A} &= Np_{\ell,A},\\
C_{-\ell,B} &= Np_{-\ell,B},\\
C_{\ell,A,-\ell,B} &= Np_{\ell,A,-\ell,B},
\end{align}
where  $N$  is the number of  photon pairs created by consecutive pump pulses during the measurement period.
For the coincidence counts $C_{\ell,A,-\ell,B}$ in the last equation it is  assumed that photon loss in beam $A$  and beam $B$ are independent. 
Note that  $p_{\ell,A,-\ell,B}/p_{\ell,A} = \eta^{(1)}_{-\ell,B}/2$ and hence one can calculate the efficiencies as:
\begin{align}
\eta^{(1)}_{-\ell,B} &= 2\frac{C_{\ell,A,-\ell,B}}{C_{\ell,A}},\\
\eta^{(1)}_{\ell,A} &= 2\frac{C_{\ell,A,-\ell,B}}{C_{-\ell,B}}.
\end{align}

\begin{table}
\[ \begin{array}{ccc}
\hline
\hline
\mbox{Basis vectors}& \mbox{Detector}~ A & \mbox{Detector}~ B\\
\hline
1 & 0.01504 & 0.02145\\ 
2 & 0.01517 & 0.02106\\
3 & 0.00536 & 0.00886\\
4 & 0.00503 & 0.00727\\
5 & 0.00508 & 0.00787\\
6 & 0.00556 & 0.00874\\
\hline
\end{array}
\]
\caption{Detection efficiencies for different detectors projecting on different bases vectors. Here the first two vectors belong to the $\sigma_z$ basis, the following two to the $\sigma_x$ basis, and the last two to the $\sigma_y$ basis.}\label{tab-1}
\end{table}

For the SLM-filter setting   $(\ket{\ell} \pm \ket{-\ell})/\sqrt{2}$ which is a superposition of $\pm \ell$ OAM modes, the corresponding creation operators read $a_{\pm}^\dagger \equiv (a_{\ell,A}^\dagger \pm a_{-\ell,A}^\dagger)/\sqrt{2}$. Thus, we can represent $a_{\ell,A}^\dagger$ and  $a_{-\ell,A}^\dagger$ in terms of $a_\pm^\dagger$ as:
\begin{align}
a_{\pm\ell}^\dagger = {a_{+,A}^\dagger \pm a_{-,A}^\dagger\over \sqrt{2}}.\label{eqn-14}
\end{align}
Substituting Eq.~\eqref{eqn-14} in Eq.~\eqref{eqn-21} we obtain:
\begin{align}
  \ket{\psi} &= \frac{\alpha}{4}\sum_{\ell =0}^\infty \sqrt{\chi_\ell}\left(\sqrt{2}\frac{(a_{+,A}^\dagger)^2}{\sqrt{2}} -  \sqrt{2}\frac{(a_{-,A}^\dagger)^2}{\sqrt{2}} \right.\nonumber\\
&\qquad\qquad \left.+ 2a_{+,A}^\dagger a_{+,B}^\dagger - 2a_{-,B}^\dagger a_{-,A}^\dagger \right.\nonumber\\
&\qquad\qquad \left.+ \sqrt{2}\frac{(a_{+,B}^\dagger)^2}{\sqrt{2}} -  \sqrt{2}\frac{(a_{-,B}^\dagger)^2 }{\sqrt{2}}\right)\ket{0}.
\end{align}
Thus, the probability of a click in detector $A$ for the SLM setting $+$ amounts to $p_{+,A} = \eta^{(1)}_{+,A}|\alpha|^2\chi_\ell/2$ while  the coincidence probability for the SLM setting $+$ in the detector $A$ and the detector $B$ reads $p_{+,A,+,B} = \eta^{(1)}_{+,A}\eta^{(1)}_{+,B}|\alpha|^2\chi_\ell/4$. The observed number of clicks are related to the expected detection counts as:
\begin{align}
C_{+,A} &= Np_{+,A},\\
C_{+,B} &= Np_{+,B},\\
C_{+,A,+,B} &= Np_{+,A,+,B},
\end{align}
Since  $p_{+,A,+,B}/p_{+,A} = \eta^{(1)}_{+,B}/2$, if follows for the efficiencies that:
\begin{align}
\eta^{(1)}_{+,B} &= 2\frac{C_{+,A,+,B}}{C_{+,A}},\\
\eta^{(1)}_{+,A} &= 2\frac{C_{+,A,+,B}}{C_{+,B}}.
\end{align}
Similarly for SLM settings $(\ket{\ell} \pm i \ket{-\ell})/\sqrt{2}$ the state $\ket{\psi}$ can be rewritten as:
\begin{align}
  \ket{\psi} &= \frac{\alpha}{4}\sum_{\ell =0}^\infty \sqrt{\chi_\ell}(\sqrt{2}\frac{(a_{+y,A}^\dagger)^2}{\sqrt{2}} +  \sqrt{2}\frac{(a_{-y,A}^\dagger)^2}{\sqrt{2}} \nonumber\\
&\qquad\qquad + 2a_{+y,A}^\dagger a_{+y,B}^\dagger + 2a_{-y,B}^\dagger a_{-y,A}^\dagger \nonumber\\
&\qquad\qquad + \sqrt{2}\frac{(a_{+y,B}^\dagger)^2}{\sqrt{2}} +  \sqrt{2}\frac{(a_{-y,B}^\dagger)^2 }{\sqrt{2}})\ket{0},
\end{align}
where
\begin{align}
a_{\pm y,A}^\dagger &= \frac{a_{\ell,A}^\dagger \pm i a_{-\ell,A}^\dagger}{\sqrt{2}}.
\end{align}
Thus, the relation for the efficiencies in this filter setting is obtained as:
\begin{align}
\eta^{(1)}_{+y,B} &= 2\frac{C_{+y,A,+y,B}}{C_{+y,A}},\\
\eta^{(1)}_{+y,A} &= 2\frac{C_{+y,A,+y,B}}{C_{+y,B}}.
\end{align}

Using the expressions derived above, we calculated the detection efficiencies for the case of a two-level system for different SLM settings (cp. Table~\ref{tab-1}). We found that even though the detection efficiencies vary for different bases, the fluctuation in the values is very small for all the basis vectors within each basis which proves the claim for qubits. 

Furthermore, this method can be used to show that the detection efficiencies are independent of the basis vectors within each basis, regardless of the dimension. However, let us point out that the analysis of our  measurement data indicated an anomaly for the detection efficiency for the OAM value $\ell = 0$, which is different from the other values of OAM. Although not so important in the present context, this case has to be investigated  more carefully when it comes to actual key transmission and will be the subject of future work.
\end{document}